\documentclass[aip,jmp,a4paper,12pt]{revtex4-1}

\usepackage[english]{babel}
\usepackage[toc,page]{appendix}
\usepackage{amsmath,amsthm,color}
\usepackage{amsfonts}
\usepackage{amssymb}

\def\ad{\text{ad\,}}
\def\diag{\mbox{diag\,}}
\def\openone{\leavevmode\hbox{\small1\kern-3.3pt\normalsize1}}
\def\bbbz{\mathbb{Z}}
\def\bbbd{\mathbb{D}}

\begin{document}

\title{\bf Integrable equations and recursion operators related to the affine Lie algebras $A^{(1)}_{r}$}

\author{V. S. Gerdjikov}
\affiliation{Institute of Nuclear Research and Nuclear Energy, Bulgarian Academy of Sciences,
72 Tsarigradsko chausee, Sofia 1784, Bulgaria}
\author{D. M. Mladenov}
\affiliation{Theoretical Physics Department, Faculty of Physics, Sofia University "St. Kliment Ohridski",
5 James Bourchier Blvd, 1164 Sofia, Bulgaria}
\author{A. A. Stefanov}
\affiliation{Theoretical Physics Department, Faculty of Physics, Sofia University "St. Kliment Ohridski",
5 James Bourchier Blvd, 1164 Sofia, Bulgaria}
\author{S. K. Varbev}
\affiliation{Theoretical Physics Department, Faculty of Physics, Sofia University "St. Kliment Ohridski",
5 James Bourchier Blvd, 1164 Sofia, Bulgaria}

\begin{abstract}
We have derived a  family of equations related to the untwisted affine Lie algebras $A^{(1)}_{r}$ using a
Coxeter $\mathbb{Z}_{r+1}$ reduction.
They represent the third member of the hierarchy of soliton equations related to the algebra.
We also give some particular examples and impose additional reductions.
\end{abstract}

\maketitle

\section{Introduction}

The main idea of the inverse scattering method (ISM) was formulated in the seminal paper  by Gardner, Greene, Kruskal and Miura \cite{GGKM}
for the Korteveg--de Vries (KdV) equation.
This method was formulated in algebraic form by Lax \cite{Lax}.
A few years later two other pioneering papers by Zakharov and Shabat \cite{ZaSha1} and Zakharov and Faddeev
\cite{ZaFa} allowed one to realize that:
i) along with the KdV equation, the non-linear Schr\"{o}dinger (NLS) equation
can also be solved by the ISM;
ii) the KdV equation is a completely integrable infinite dimensional Hamiltonian system.
Zakharov and Shabat also developed the so-called dressing method \cite{ZaSha2} that is now one of the most effective methods for calculating
soliton solutions of these equations.
Today this method bridges several areas of modern mathematics, mathematical and applied physics, see \cite{CalDeg,FaTa,ZMNP,GYaV*08}.

Another important idea which was proposed by Ablowitz, Kaup, Newell and Segur \cite{AKNS} is that the ISM can be understood
as a generalized Fourier transform.
This idea was rigorously proved \cite{Kaup,GeKh1} and extended to a large class
of Lax operators \cite{GeKu*81,Ge4,SIAM*14}.

The next important milestone of the soliton theory was laid by the famous papers of Mikhailov \cite{Mikh} and Drinfeld and Sokolov \cite{DriSok} in the early 1980-ies.
The first one not only discovered the important class of 2-dimensional Lorentz invariant field theories: the 2-dimensional Toda field theories, but also introduced
the group of reductions as an important tool for discovering new integrable systems. The second paper \cite{DriSok} established
the close relation between the Lax representations and the Kac-Moody algebras.

It is not possible to explain in a few words the deep ideas developed in all these papers and their consequences.
However, outlining the milestones of the soliton theory, they also raised a number of `minor' problems.

This is the first of a series of papers in which we plan to detail one by one these ideas for the class of soliton equations with deep reductions.
We will focus mainly on the mKdV-type equations but obviously these results
can easily be extended to any of the members of the hierarchy related to the given Lax operator.

The first of these problems  is to construct an explicit realization (or the simplest matrix representation) of the Kac-Moody algebras relevant to the  Lax pairs.
Here we start with the class of (untwisted) height $1$ Kac-Moody algebras.

This work can be considered as an extension of our recent publications \cite{GMSV1, GMSV2, GMSV3}.
We extensively use the technique introduced in the classical paper by Drinfeld and Sokolov \cite{DriSok} to implement:
 i) the graded algebras needed for the Lax representation and the automorphisms of Coxeter type introducing the corresponding grading;
ii) the recursion operators and the Hamiltonian formulation of the integrable systems.
The completion of all these tasks is still a challenge which, we believe, must be met.

As we already mentioned above, the general theory of the nonlinear evolution equations (NLEE) allowing Lax representation is well developed
\cite{DriSok,CalDeg,FaTa,ZMNP,AKNS,Ge4,GeKu*81,67}.
This paper deals with NLEE that allow Lax representation with deep reductions \cite{Mikh}.
This means that they can be written as the commutativity condition of two differential operators of the type
\begin{eqnarray}
&&
L\psi \equiv i\frac{\partial \psi}{ \partial x } + U(x,t,\lambda)\psi = 0, \label{1a}\\
&&
M\psi \equiv i\frac{\partial \psi}{ \partial t } + V(x,t,\lambda)\psi = \psi \Gamma(\lambda),
\label{1b}
\end{eqnarray}
where $U(x,t,\lambda)$, $V(x,t,\lambda)$ and $\Gamma(\lambda)$ are some polynomials of the spectral parameter $\lambda$
with coefficients taking values in the algebra $A_r \simeq sl(r+1)$. The Coxeter number of $sl(r+1)$ is equal to $r+1$.
That is why we request also that the Lax pair (\ref{1a}) and (\ref{1b}) possesses $\bbbz_{r+1} $-reduction group.
This means that we impose on (\ref{1a}) and (\ref{1b}) a $\bbbz_{r+1}$-reduction by \cite{Mikh}
\begin{equation}\label{2}
C_1 U(x,t,\lambda) C_1^{-1} = U(x,t,\omega\lambda),\quad C_1 V(x,t,\lambda) C_1^{-1} = V(x,t,\omega\lambda),
\end{equation}
where $C_1^{r+1} =\openone  $ is a Coxeter automorphism  of the algebra $A^{(1)}_{r}$ and $\omega =e^{\frac{2\pi i}{r+1}} $.

Some of the  relevant NLEE related to this class of Lax operators have already been studied. The most famous of
them are the 2-dimensional Toda field theories \cite{Mikh}
\begin{equation}\label{eq:2tft}\begin{split}
\frac{\partial^2 w_j }{ \partial x \partial t} = e^{w_{j+1}-w_j} - e^{w_{j}-w_{j-1}} , \qquad j=1,\dots , r+1,
\end{split}\end{equation}
where the indices $j\pm 1$ are considered modulo $r+1$, i.e. $w_0\simeq w_{r+1}$ and $w_{r+2}\simeq w_{1}$.

Another class of such equations is known as a generalization of the derivative nonlinear Schr\"{o}dinger (DNLS) equation \cite{VSG-88,VG-13}
(see also \cite{Mikh,KauNew})
\begin{equation}\begin{split}\label{3}
i \frac{\partial \psi _{k}}{ \partial t } + \gamma \frac{\partial }{ \partial x } \left( \cot \left(\frac{\pi k}{ r+1 }\right) \cdot \psi _{k,x} +
i \sum_{p=1}^{r} \psi _p \psi _{k-p} \right) = 0,
\end{split}\end{equation}
where $ k=1,2,\dots, r$, $\gamma  $ is a constant, again the index $k-p $ should be understood modulo $r+1$ and $\psi _0 = \psi _{r+1} =0 $.

The paper is organized as follows. Section 2 contains preliminaries necessary to derive the NLEE.
In particular we provide a convenient basis for $A^{(1)}_{r}$ Kac-Moody algebra which is compatible
with the $\bbbz_{r+1}$-reduction. In Section 3 we derive the recursion relations following from the Lax
representation. Here we assume that the $M$ operator is polynomial in the spectral parameter $\lambda$.
Following the ideas of AKNS \cite{AKNS} we show that these recursion relations can be solved by
the recursion operators
(see \cite{GYaV*08,Ge4,VG-Ya-13,Yan2013a}) which are obtained in factorized form \cite{Yan2013a,YanVi*2012b,SIAM*14}.
We also derive the mKdV equations for any $r$.
In section 4 we explain the idea of the ISM for this particular choice of the Lax operator $L$
and derive the evolution equations for scattering matrix of $L$. In Section 5 we give particular examples
of DNLS and mKdV-type equations together with their Hamiltonians.
In Section 6 it is shown that one can impose additional $\bbbz_2$-reductions  on these  equations.
Section 7 is devoted to the generating functionals of the class of $M$-operators compatible with $L$.
We end the paper with some concluding remarks.

\section{Preliminaries }

Let us start with two different choices $C_1$ and $\tilde{ C}_1$ for the Coxeter  automorphism
for the algebra $\mathfrak{g}\simeq \mathfrak{sl}(r+1)$.
In what follows we assume that the reader is familiar with the basic concepts of the simple and affine Lie algebras,
see for example \cite{Kac,Cart,Helg}.
Each of these choices satisfies $C_1^{r+1}=\openone$ and  $\tilde{ C}_1^{r+1}=\openone$, and each of these automorphisms induces a  grading in $\mathfrak{g}$
\begin{equation}\label{eq:grad1}\begin{aligned}
\mathfrak{g} &= \mathop{\oplus}\limits_{k=0}^{r} \mathfrak{g}^{(k)}, &\qquad
\tilde{\mathfrak{g}} &= \mathop{\oplus}\limits_{s=0}^{r} \tilde{\mathfrak{g}}_{s}\,.
\end{aligned}\end{equation}
Here the linear subspaces are such that
\begin{equation}\label{eq:grad2}
C_1 X C_1^{-1} = \omega^{-k} X, \qquad  \tilde{ C}_1 Y \tilde{ C}_1^{-1} = \omega^{-s} Y,
\end{equation}
where $X \in \mathfrak{g}^{(k)}, Y \in \tilde{ \mathfrak{g}}_{s}$ and $\omega= e^{2\pi i/(r+1)}$.

Each of the gradings satisfies
\begin{equation}\label{eq:grad3}\begin{split}
[ \mathfrak{g}^{(k)}, \mathfrak{g}^{(m)}] \in \mathfrak{g}^{(k+m)}, \qquad
[ \tilde{ \mathfrak{g}}_{s}, \tilde{ \mathfrak{g}}_{p}] \in \tilde{ \mathfrak{g}}_{s+p},
\end{split}\end{equation}
where $(k+m)$ and $(s+p)$ must be understood modulo $(r+1)$.

In what follows we will specify the choice of the automorphisms by
\begin{equation}\label{eq:grad4}\begin{split}
C_1 = \sum_{p=1}^{r+1} E_{p, p+1}=J_{1}^{(0)}, \qquad \tilde{C}_1 = \sum_{p=1}^{r+1} \omega^{p-1} E_{p,p}=J_{0}^{(1)},
\end{split}\end{equation}
where the $(r+1)\times(r+1$) matrices $E_{km}$ are defined by $(E_{km})_{sp} = \delta_{ks}\delta_{mp}$.

Further we will use a convenient basis in the affine Lie algebra $A^{(1)}_{r}$ which satisfies both  of the above gradings
(see  \cite{Helg,Kac,Cart})
\begin{equation}\label{5}
J_s^{(k)} = \sum_{j=1}^{r+1} \omega ^{kj} E_{j,j+s}.
\end{equation}
The elements  of this basis satisfy the commutation relations
\begin{equation}\label{6}
\left[ J_s^{(k)} , J_l^{(m)} \right] = \left(\omega ^{ms} - \omega ^{kl} \right) J_{s+l}^{(k+m)}.
\end{equation}
Besides it is easy to check that
\begin{equation}
\label{eq:C1}
\begin{split}
 C^{-1}_{1} J_s^{(k)} C_{1} = \omega^{-k} J_s^{(k)}, \qquad \tilde{ C}^{-1}_{1} J_s^{(k)}
 \tilde{ C}_{1} = \omega^{-s} J_s^{(k)}
\end{split}
\end{equation}
and
\begin{equation}
\begin{split}
J_{s}^{(k)}J_{p}^{(m)} = \omega^{sm} J_{s+p}^{(k+m)}, \qquad (J_{s}^{(k)})^{-1}=(J_{s}^{(k)})^{\dagger}.
\end{split}
\end{equation}

Using this we can specify the bases in each of the linear subspaces as follows
\begin{equation}\label{eq:grad7}\begin{split}
\mathfrak{g}^{(k)} \equiv \mbox{span} \{ J_s^{(k)}, \quad s=1, \dots, r+1\}, \qquad
\tilde{ \mathfrak{g}}_{s} \equiv \mbox{span} \{ J_s^{(k)}, \quad k=1, \dots, r+1\}.	
\end{split}\end{equation}

 The realization of the Coxeter automorphism by $C_{1}$ corresponds to choosing it as a Weyl group element
 $C_1=S_{\alpha_1} \dots S_{\alpha_r}$, where $\alpha_k$ are the simple roots of ${\mathfrak{sl}}(r+1)$ and $S_\alpha$
 is the Weyl reflection with respect to the root $\alpha$.
 In the other realization $\tilde{C}_1$ is an element of the Cartan subgroup of $\mathfrak{sl}(r+1)$.
 Then $\mathfrak{g}^{(0)} \equiv \mathfrak{h}$ is the Cartan subalgebra  of $\mathfrak{sl}(r+1)$.
 Both realizations are equivalent, i.e. there exists a similarity transformation which takes $C_{1}$ into $\tilde{C}_1$.
 In the first realization each of the linear subspaces $\mathfrak{g}^{(k)}$ (with the exception of ${\mathfrak{g}}^{(0)}$) has a one-dimensional section with
 the Cartan subalgebra, i.e.
\begin{equation}
\label{eq:grad8}
\begin{split}
\mathfrak{g}^{(s)} \cap \mathfrak{h} \equiv \alpha_{s} J_{0}^{(s)},
\end{split}
\end{equation}
where $\alpha_{s}$ is an arbitrary constant.

\section{Lax Representations and Integrable Equations }

\subsection{Lax Representations}
We start with a Lax pair that is polynomial in the spectral parameter $\lambda$
\begin{eqnarray}
&&
\label{LaxPair1}
L = i \partial_x + Q(x,t) - \lambda J, \\
&&
\label{LaxPair2}
M = i \partial_t  + \sum_{k=0}^{m-1} \lambda^k V^{(k)}(x,t) - \lambda^{m} K,
\end{eqnarray}
where
\begin{equation}
\label{8}
Q(x,t) \in \mathfrak{g}^{(0)}, \quad V^{(k)}(x,t) \in \mathfrak{g}^{(k)}, \quad K \in \mathfrak{g}^{(m)}, \quad J\in \mathfrak{g}^{(1)}.
\end{equation}
Here $J$ and $K$ are some properly chosen constant matrices.
In order to get a set of mKdV equations we have chosen $V(x,t,\lambda)$ to be a cubic polynomial of $\lambda$.

The Lax pair must commute, i.e.
\begin{equation}
\label{LPC}
\big[L,\, M \big]=0
\end{equation}
for every $\lambda$, which will lead to a set of recursion relations.
Solving them we will get explicit expressions for $V^{(k)}(x,t)$ in terms of $Q(x,t)$ and
finally will obtain the equations as constraints for the potential $Q(x,t)$.

Here we adapt the theory of the recursion operators \cite{GYaV*08,Ge4,VG-Ya-13,Yan2013a}
to the special choices of the Lax pair (\ref{LaxPair1}) and (\ref{LaxPair2}).
For simplicity we assume that $m$ in (\ref{LaxPair1}) and (\ref{LaxPair2}) is $m \leq (r+1)$ and $m$ is an exponent of $\mathfrak{g}$. If $m>(r+1)$ then $m$ should be understood modulo $r+1$ and if $m$ is not an exponent the equations will be trivial.
The commutativity condition \eqref{LPC} implies the following recursion relations
\begin{equation}
\label{Recurrence1}
\begin{aligned}
&\lambda^{m+1}: & \big[ J, K \big]&=0, \\
&\lambda^{m}:  & \big[ J,V^{(m-1)}(x,t) \big]+ \big[ Q(x,t), K \big]&=0, \\
&\lambda^{s}: &  i\frac{\partial V^{(s)}}{\partial x}+ \big[Q(x,t),V^{(s)}(x,t) \big]- \big[ J,V^{(s-1)}(x,t) \big] &=0, \\
&\lambda^{0}: &  - i\frac{\partial Q}{\partial t}+  i\frac{\partial V^{(0)}}{\partial x}+  \big[ Q(x,t),V^{(0)}(x,t) \big]&=0. \\
\end{aligned}
\end{equation}

By $\mbox{ad}_{J}$ we mean the linear operator defined by
\begin{equation}\label{7}
\mbox{ad}_{J}(X) = \big[J,\, X \big].
\end{equation}
This operator has a kernel and can only be inverted on its image.
We denote that inverse by $\mbox{ad}_J^{-1}$. From its spectral properties it follows that $\mbox{ad}_J^{-1}$
can be expressed as a polynomial in $\mbox{ad}_J$, see \cite{PLA110}.

Each element splits into "orthogonal" and "parallel" parts
\begin{equation}
\begin{aligned}
V^{(s)}(x,t)&=V^{(s)}_{\bot}(x,t)+V^{(s)}_{\|}(x,t), \quad \mbox{ad}_J \left(V^{(s)}_{\|}(x,t) \right)=0, \\
V^{(s)}_{\|}(x,t)&=
 \begin{cases}
0 &\mbox{if } s \mbox{ is equal to $0$,} \\
c_{s}^{-1} J^s \left< V^{(s)}, J^{r+1-s} \right> &\mbox{if } s\ge 1 \mbox{ is an exponent,}
\end{cases}
\end{aligned}
\end{equation}
where $c_{s} =  \left< J^{s}, J^{r+1-s} \right>=r+1$.
Here with $\big\langle . \,, . \big\rangle$ is denoted the Killing-Cartan form on $\mathfrak{g}$.

From \eqref{Recurrence1} it can be obtained that
\begin{equation}
\begin{aligned}
V^{(s-1)}_{\bot}(x,t) &= \mbox{ad}_{J}^{-1}\left( i\frac{\partial V^{(s)}}{\partial x}+ \big[ Q(x,t),V^{(s)}_{\bot}(x,t) \big] + \big[ Q(x,t),V^{(s)}_{\|}(x,t) \big]  \right), \\
i  \frac{\partial V^{(s)}_{\|}}{\partial x} &=- \big[ Q(x,t),V^{(s)}_{\bot}(x,t) \big]_{\|}.
\end{aligned}
\end{equation}
After integration of the second equation we get
\begin{equation}\begin{aligned}
 V^{(s)}_{\|}(x,t)&=i(\partial_x)^{-1}\left(\left[Q(x,t),V^{s}_{\bot}(x,t) \right]_{\|} \right) \\
 &= \frac{i}{r+1} J^{s} (\partial_x)^{-1} \left< \left[Q(x,t),V^{s}_{\bot}(x,t) \right], J^{r+1-s} \right>,
\end{aligned}
\end{equation}
where $(\partial_x)^{-1}=\int_{- \infty}^{x} dy$ and we have set any integration constants to be equal to zero.
Thus the formal solution of the recurrent relations takes the form
\begin{equation}
V^{(s)}_{\bot}(x,t) =  \Lambda_{s+1} V^{(s+1)}_{\bot}(x,t).
\label{RecOp}
\end{equation}
If $s=0$ (mod $(r+1)$) then
\begin{equation}\label{Lambda-s} \begin{aligned}
\Lambda_{0} X &= \mbox{ad}_{J}^{-1}\left( i\frac{\partial X}{\partial x}+ \big[ Q(x,t), X \big] \right),
\end{aligned}\end{equation}
otherwise
\begin{equation}
\begin{aligned}
\Lambda_{s} X &= \mbox{ad}_{J}^{-1}\left( i\frac{\partial X}{\partial x}+ \big[ Q(x,t), X \big] 
 +\frac{i}{r+1} \big[ Q(x,t), J^s \big](\partial_x)^{-1} \left< \left[Q(x,t), X \right], J^{r+1-s} \right>  \right).
\end{aligned}\end{equation}
We remind that  $\ad_J$ is defined by (\ref{7}) where $X$ is any element of the Lie algebra. The operator $\ad_J$ has as a
kernel the Cartan subalgebra $\mathfrak{h}$ of $\mathfrak{g}$. Therefore it can be inverted only on $\mathfrak{g}\slash \mathfrak{h}$.
In matrix form  $\ad_J^{-1}$ can be calculated as follows:
\begin{equation}\label{eq:adJ-1}\begin{split}
\left( \ad_J^{-1} X^\perp \right)_{mn} = \frac{X_{mn} }{J_m - J_n}, \qquad m\neq n.
\end{split}\end{equation}
In particular, if
\begin{equation}
J = \diag ( \omega, \omega^2 ,\dots , \omega^r , 1 ) ,  \qquad
J^{m}= \diag \left( \omega^m, \omega^{2m}, \dots, \omega^{mr}, 1 \right)
\end{equation}
then
\begin{equation}\label{eq:adJ-1'}\begin{split}
\left( \ad_J^{-1} [J^m ,Q] \right)_{pn} = \frac{\omega^{mp} - \omega^{mn} }{\omega^{p} - \omega^{n}} Q_{pn}.
\end{split}\end{equation}

The elements of the potential of $M$ are given by
\begin{equation}
\label{eq:Vm}
\begin{split}
V^{(m-1)} &= \ad_{J^{-1}} [J^{m},Q], \\
V^{(m-2)}_\perp  &= \Lambda_{m-1} \ad_{J^{-1}} [J^{m},Q], \\
V^{(m-3)}_\perp  &= \Lambda_{m-2} \Lambda{_{m-1}} \ad_{J^{-1}} [J^{m},Q], \\
& \vdots \\
V^{(0)}_\perp  &= \Lambda_{1}  \cdots \Lambda_{m-2} \Lambda_{m-1} \ad_{J^{-1}} [J^{m},Q].
\end{split}
\end{equation}
Thus we obtain the $M$ operator for the NLEE with dispersion relation $\lambda^{m}$.

The corresponding NLEE can be written as
\begin{equation}
i \mbox{ ad}_{J}^{-1} \frac{\partial Q}{\partial t} - \Lambda_{0} V^{(0)}=0
\end{equation}
and $\Lambda_{0}$ is given by (\ref{Lambda-s}).

\subsection{Derivation of the equations}

Let us consider the Lax pair (\ref{LaxPair1}) and (\ref{LaxPair2}) where $K \sim J^m$ and $m=3$
\begin{equation}\label{eq:L}\begin{split}
L\psi \equiv  \biggl( i\frac{\partial}{\partial x}+Q(x,t)-\lambda J \biggr) \psi=0,
\end{split}\end{equation}
\begin{equation}\label{eq:M}\begin{split}
M\psi &\equiv  \biggl( i\frac{\partial}{\partial t}+V_{0}(x,t)+\lambda V_{1}(x,t)+\lambda^{2}V_{2}(x,t)-\lambda^{3} K \biggr) \psi  \\
&= -\lambda^{3} \psi K.
\end{split}\end{equation}

The zero-curvature condition $[L\,,M] = 0$ leads to a polynomial of fourth order in $\lambda$ which has to be identically zero.
The analog of equations (\ref{Recurrence1}) leads to:
\begin{eqnarray}
&&
\label{10}
\lambda^{4}: [J,K]=0\,, \\
&&
\label{11}
\lambda^{3}: [J,V_{2}]=[K,Q]\,,\\
&&
\label{12}
\lambda^{2}:  [J,V_{1}]=[Q,V_{2}]+ i\frac{\partial V_{2}}{\partial x}\,,\\
&&
\label{13}
\lambda^{1}:  [J,V_{0}]=[Q,V_{1}]+ i\frac{\partial V_{1}}{\partial x}\,, \\
&&
\label{14}
\lambda^{0}:  i\frac{\partial Q}{\partial t}=[Q,V_{0}]+ i\frac{\partial V_{0}}{\partial x}\,.
\end{eqnarray}
Here $Q$ and $J$ have the form
\begin{equation}
\label{15}
Q(x,t)=\sum_{j=1}^{r} q_{j}(x,t)J_{j}^{(0)}\,, \qquad
J=aJ_{0}^{(1)}
\end{equation}
where $a$ is a constant.
From (\ref{10}) it follows that
\begin{equation}
\label{17}
K=bJ_{0}^{(3)},
\end{equation}
where $b$ is another constant.
For $V_{2}(x,t)$ we have
\begin{equation}
\label{18}
V_{2}(x,t)=\sum_{k=1}^{r+1}v_{k}^{(2)}(x,t)J_{k}^{(2)}.
\end{equation}
 From equation (\ref{11}) we find that
\begin{equation}
\label{19}
v_{j}^{(2)}=\frac{b}{a}(\omega^{2j}+\omega^{j}+1) q_{j},
\end{equation}
where $j$ runs from $1$ to $r$.
To obtain $v_{r+1}^{(2)}$ we have to take into account the diagonal part of (\ref{12}).
This leads to
\begin{equation}
\label{20}
i\frac{\partial v_{r+1}^{(2)}}{\partial x}=0
\end{equation}
with a solution
\begin{equation}
\label{21}
v_{r+1}^{(2)}=0
\end{equation}
up to a constant.

For $V_{1}(x,t)$ we find that
\begin{equation}
\label{22}
V_{1}(x,t)=\sum_{l=1}^{r+1}v_{l}^{(1)}(x,t)J_{l}^{(1)}.
\end{equation}
From the part that does not commute identically with $J$ in (\ref{12}) we obtain
\begin{equation}\label{23}
\begin{split}
v_{l}^{(1)}=\frac{b}{a^{2}} \sum_{j+k=l}^{r}\frac{\omega^{2l}+\omega^{2j+k}-\omega^{k}-1}{1-\omega^{l}} q_{j} q_{k}+i\frac{b}{a^{2}}\left( \frac{\omega^{2l}+\omega^{l}+1}{1-\omega^{l}}\right) \frac{\partial q_{l}}{\partial x},
\end{split}
\end{equation}
where $l$ is running from $1$ to $r$.
To obtain $v_{r+1}^{(1)}$, again we have to take into account the diagonal part of (\ref{13}).
This leads to
\begin{equation}
\label{24}
\frac{\partial v_{r+1}^{(1)}}{\partial x}=-\frac{b}{a^{2}} \sum_{j+l=1}^{r+1}
\left( \cos{ \left( \frac{2\pi j}{r+1} \right) }+\frac{1}{2}\right) \frac{\partial}{\partial x}(q_{j} q_{l})
\end{equation}
with a solution
\begin{equation}
\label{25}
v_{r+1}^{(1)}=-\frac{b}{a^{2}} \sum_{j+l=1}^{r+1}\left( \cos{ \biggl( \frac{2\pi j}{r+1} \biggr) }+\frac{1}{2}\right) q_{j} q_{l}
\end{equation}
up to a constant.

From the orthogonal part of (\ref{13}) for $V_{0}(x,t)$ we get
\begin{equation}
\label{26}
V_{0}(x,t)=\sum_{m=1}^{r}v_{m}^{(0)}(x,t)J_{m}^{(0)}.
\end{equation}
Solving for the coefficients we find that
\begin{equation}
\begin{aligned}\label{27}
v_{j}^{(0)}&= \frac{b}{a^{3}} \Bigg(  \sum_{l+m=j}^{r}\sum_{i+k=l}^{r}\left(\frac{\sin{ \left( \frac{\pi(j-2k)}{r+1} \right) }-\sin{\left( \frac{\pi(j-2m)}{r+1}\right)}
  }{\sin{ \left( \frac{\pi j}{r+1} \right) }}+1\right)(q_{i} q_{k} q_{m})  \\
&+ \sum_{k=1}^{r}\left( \cos{ \left( \frac{2\pi k}{r+1} \right) }+\frac{1}{2}\right)( q_{k} q_{r+1-k} q_{j})
+  \frac{  4\cos^{2}{ \left( \frac{\pi j}{r+1} \right) -1}}{4\sin^{2}{ \left( \frac{\pi j}{r+1} \right) }} \frac{\partial^{2} q_{j}}{\partial x^{2}}\\
  &+\frac{3}{2}\sum_{l+m=j}^{r}\cot \left( {\frac{\pi l}{r+1}}\right)\frac{\partial q_{l}}{\partial x} q_{m}+\frac{3}{4}\cot \left( {\frac{\pi j}{r+1}}\right) \sum_{k+l=j}^{r}\frac{\partial}{\partial x}(q_{k} q_{l}) \Bigg).
\end{aligned}
\end{equation}
From here we obtain the set of mKdV-type equations
\begin{equation}
\begin{aligned}\label{eq:30'}
\frac{\partial \psi_{j}}{\partial t} &= \frac{b}{a^3} \frac{\partial }{ \partial x } \Bigg( \frac{ s_{3j}}{4s_j^3} \frac{\partial^2 q_{j} }{ \partial x^2 } + \frac{3}{4} \sum_{k+l=j}^{} \left(  \frac{c_l }{s_l} \frac{\partial q_{l}}{ \partial x } q_{k} + \frac{c_k }{s_k}
\frac{\partial q_{k}}{ \partial x } q_{l}+ \frac{c_j}{s_j} \frac{\partial (q_{k}q_{l})}{ \partial x } \right)   \\
 &+ \frac{1}{2}\sum_{k=1}^{r} (4c_k^2 - 1 ) q_{k} q_{r+1-k} q_{j} + \sum_{ l+m=j}^{} \sum_{ i+k=l}^{}
 \left( \frac{ 2s_{m-k}c_{j-k-m}}{s_j} +1 \right) q_{i} q_{k} q_{m} \Bigg)
\end{aligned}
\end{equation}
where we have used the notations
\begin{equation}\label{eq:not}\begin{split}
c_j = \cos \frac{ \pi j}{r+1}, \qquad  s_j = \sin \frac{ \pi j}{r+1}.
\end{split}\end{equation}

\section{Solving the NLEE through the Inverse Scattering method }

\subsection{The inverse scattering method}

We can solve this system of equations via the Inverse scattering method (ISM).
Let's $q_{0}$ is some solution of the mKdV equations at time $t=0$ i.e.
 \begin{equation}
 q_0(x)= Q(x,t=0)  .
 \end{equation}
We can divide the method in three steps.

{\bf Step I:} First we solve the direct scattering problem, which means that if we know some $t$-independent solution
$q_{0}(x)$ we have to construct the scattering matrix $T(\lambda,0)$ which can be done using the so called Jost solutions.
The Jost solutions are given by
\begin{equation}\label{eq:55}\begin{split}
 \lim _{x\to -\infty} \phi_-(x,\lambda) e^{i\lambda Jx} &=\openone, \qquad \lim _{x\to \infty} \phi_+(x,\lambda) e^{i\lambda Jx} =\openone, \\
L\phi_-(x,\lambda) &=0, \qquad L\phi_+(x,\lambda) =0,
\end{split}\end{equation}
and the scattering matrix is given in terms of these solutions
\[ T(\lambda,0)  = \phi_+^{-1}(x,\lambda) \phi_-(x,\lambda). \]

{\bf Step II:} If we know the scattering matrix at time $t=0$ we can calculate it at any time.
This step will be considered especially in the next subsection.

{\bf Step III:} Now we have to solve the inverse scattering problem, which means that given the scattering matrix at time $t\neq0$
we have to construct the potential $Q(x,t)$ for $t>0$. \\
It means that we have to solve the Gelfand-Levitan-Marchenko equation. In our case this is integral equation of Fredholm type.
We have two options. To solve it directly which is very complicated task or to reduce this integral
equation to Riemann-Hilbert problem.

With this three steps we reduce the nonlinear Cauchy problem to a sequence of three linear Cauchy problems
each of them having an unique solution.

\subsection{The evolution of the scattering matrix}

Both Jost solutions $\phi_\pm(x,t,\lambda)$ satisfy the equations (\ref{1a}) and (\ref{1b}).
Let us now calculate the limit
\begin{equation}
\label{eq:limM}
\begin{split}
\lim_{x \to \infty} M\phi_{+}(x,t) &=
\lim_{x\to\infty}\left[\left(i\frac{\partial}{\partial t}+V_{0}+\lambda V_{1}+\lambda^2 V_{2} \dots -\lambda^{m} K \right) e^{-i\lambda J_{0}^{(1)} x} \right]\\
&=e^{-i\lambda J_{0}^{(1)} x} \Gamma(\lambda).
\end{split}
\end{equation}
Assuming that the definitions of the Jost solutions are $t$-independent we find that
\begin{equation}
\Gamma(\lambda)=\lim_{x\to\infty}(V_{0}+\lambda V_{1}+\lambda^2 V_{2} \dots -\lambda^{m} K)=-\lambda^{m} K.
\end{equation}
Next we calculate
\begin{equation}\label{eq:Mphim}\begin{split}
\lim_{x \to\infty} M\phi_{-}(x,t) &=
\left(i\frac{\partial}{\partial t}-\Gamma(\lambda)\right) e^{-i\lambda J_{0}^{(1)} x} T(\lambda,t)\\
&= e^{-i\lambda J_{0}^{(1)} x} \left(i\frac{\partial T}{\partial t}- \Gamma(\lambda)  T(\lambda,t)\right)\\
&=e^{-i\lambda J_{0}^{(1)} x} T(\lambda,t) \Gamma(\lambda).
\end{split}\end{equation}
Therefore, if $Q(x,t)$ satisfies the mKdV equations (\ref{eq:30'}) the scattering matrix $T(\lambda,t)$
must satisfy the following linear evolution equation
\begin{equation}\label{eq:T-t}
i \frac{\partial T}{\partial t}-[\Gamma(\lambda),T(\lambda,t)]=0.
\end{equation}
In the particular case when $V_{i}=0$ we get
\begin{equation}\label{eq:T-t1}
i \frac{\partial T}{\partial t} +\lambda^{s} [K,T(\lambda,t)]=0,
\end{equation}
whose solution is
\begin{equation}
T_{ij}(\lambda,t)= e^{i\lambda^{s} (\omega^{si}-\omega^{sj})t} T_{ij}(\lambda,0).
\end{equation}
Thus $T_{ij}(\lambda,0)$ is the Cauchy data for the initial conditions of the scattering matrix.
Therefore solving the mKdV equations (\ref{eq:30'}) reduces to solving the direct and the inverse
scattering problem for the Lax operator $L$, see \cite{Mikh,VG-13,VG-Ya-13}.

\section{Examples and Hamiltonian properties}

\subsection{DNLS type equations}

Special examples of DNLS systems of equations can be found in \cite{VG-13}.
We will give some particular examples when $M$ operator is from second and third degree in $\lambda$.

Those equations admit the following Hamiltonian formulation
\begin{equation}
\frac{\partial q_{i}}{\partial t}=\frac{\partial}{\partial x} \biggl( \frac{\delta H}{\delta q_{r+1-i}} \biggr) .
\end{equation}

The first interesting nontrivial case is when $M$ is quadratic polynomial in $\lambda$ and
 $\mathfrak{g}\simeq A^{(1)}_{2}$ algebra. The potential of $L$ is given by
 \begin{equation}
\begin{split}
 U(x,t,\lambda) = \left(\begin{array}{cccc}
  0 & q_1 & q_2 \\
  q_2 & 0 & q_1 \\
  q_1 & q_2 & 0
  \end{array}\right)
  - \lambda \left(\begin{array}{cccc}
  1 & 0 & 0  \\
  0 & \omega & 0  \\
  0 & 0 & \omega^2
  \end{array}\right),
\end{split}
\end{equation}
where $\omega= e^{2\pi i/3}$.
This gives us the system of integrable nonlinear partial differential equations
\begin{equation}
\label{121}
i\frac{\partial q_{1}}{\partial t}+i\gamma\frac{\partial}{\partial x}(q_{2}^{2})+\gamma\frac{\sqrt{3}}{3}\frac{\partial^{2} q_{1}}{\partial x^{2}}=0,
\end{equation}
\begin{equation}
\label{122}
i\frac{\partial q_{2}}{\partial t}+i\gamma\frac{\partial}{\partial x}(q_{1}^{2})-\gamma\frac{\sqrt{3}}{3}\frac{\partial^{2} q_{2}}{\partial x^{2}}=0.
\end{equation}
The corresponding Hamiltonian is
\begin{equation}
H=\frac{i\gamma \sqrt{3}}{6} \biggl( q_{2}\frac{\partial q_{1}}{\partial x}-q_{1}\frac{\partial q_{2}}{\partial x} \biggr)-
\frac{\gamma}{3}(q_{1}^{3}+q_{2}^{3}).
\end{equation}

In the case of $A^{(1)}_{3}$ algebra using the potential
\begin{equation}
\begin{split}
 U(x,t,\lambda) = \left(\begin{array}{cccc}
  0 & q_1 & q_2 & q_3 \\
  q_3 & 0 & q_1 & q_2 \\
  q_2 & q_3 & 0 & q_1 \\
  q_1 & q_2 & q_3  & 0
  \end{array}\right)
  - \lambda \left(\begin{array}{cccc}
  1 & 0 & 0 & 0 \\
  0 & i & 0 & 0 \\
  0 & 0 & -1 & 0 \\
  0 & 0 & 0 & -i
  \end{array}\right),
\end{split}
\end{equation}
we obtain the system of integrable nonlinear partial differential equations
\begin{equation}
\label{123}
i\frac{\partial q_{1}}{\partial t}+2i\gamma\frac{\partial}{\partial x}(q_{2}q_{3})+\gamma\frac{\partial^{2} q_{1}}{\partial x^{2}}=0,
\end{equation}
\begin{equation}
\label{124}
i\frac{\partial q_{2}}{\partial t}+i\gamma\frac{\partial}{\partial x}(q_{1}^{2})+i\gamma\frac{\partial}{\partial x}(q_{3}^{2})=0,
\end{equation}
\begin{equation}
\label{125}
i\frac{\partial q_{3}}{\partial t}+2i\gamma\frac{\partial}{\partial x}(q_{1}q_{2})-\gamma\frac{\partial^{2} q_{3}}{\partial x^{2}}=0.
\end{equation}
The corresponding Hamiltonian is
\begin{equation}
H=\frac{i\gamma}{2} \biggl( q_{3}\frac{\partial q_{1}}{\partial x}-q_{1}\frac{\partial q_{3}}{\partial x}
+\frac{1}{2}\frac{\partial}{\partial x}(q_{2}^{2}) \biggr)-\gamma q_{2}(q_{1}^{2}+q_{3}^{2}).
\end{equation}

\subsection{Systems of equations of mKdV type}

These are equations with cubic dispersion laws, therefore the $M$-operators are also cubic polynomials in $\lambda$.

In the case of $A^{(1)}_{1}$ algebra, with the following potential
\begin{equation}
\begin{split}
 U(x,t,\lambda) = \left(\begin{array}{cccc}
  0 & q_1  \\
  q_1 & 0
 \end{array}\right)
  - \lambda \left(\begin{array}{cccc}
  1 & 0  \\
  0 & -1    \end{array}\right),
\end{split}
\end{equation}
we obtain the well-known focusing mKdV equation
\begin{equation}
\label{1mKdV}
\alpha \frac{\partial q_{1}}{\partial t}=-\frac{1}{4}\frac{\partial^{3}
q_{1}}{\partial x^{3}}-\frac{1}{2}\frac{\partial}{\partial x}(q_{1}^{3})\,,
\end{equation}
where $\alpha=\frac{a^3}{b}$.
In this case the Hamiltonian is
\begin{equation}
H=\frac{1}{8\alpha} \left( \left( \frac{\partial q_{1}}{\partial x} \right)^{2}-q_{1}^{4} \right).
\end{equation}
In the case of $A^{(1)}_{2}$ algebra we obtain a trivial system of equations
$\partial_t q_{1} = 0$ and $\partial_t q_{2} = 0$ and the corresponding Hamiltonian is bilinear with
respect to $q_{1}$ and $q_{2}$.

In the case of $A^{(1)}_{3}$ algebra the potential of the Lax operator is parameterized by
\begin{equation}
\begin{split}
 U(x,t,\lambda) = \left(\begin{array}{cccc}
  0 & q_1 & q_2 & q_3 \\
  q_3 & 0 & q_1 & q_2 \\
  q_2 & q_3 & 0 & q_1 \\
  q_1 & q_2 & q_3  & 0
  \end{array}\right)
  - \lambda \left(\begin{array}{cccc}
  1 & 0 & 0 & 0 \\
  0 & i & 0 & 0 \\
  0 & 0 & -1 & 0 \\
  0 & 0 & 0 & -i
  \end{array}\right),
\end{split}
\end{equation}
which is related to the following system of mKdV type equations
\begin{equation}
\label{eq:mk1-3}
\begin{split}
 \alpha \frac{\partial q_{1}}{\partial t} &=\frac{1}{2}\frac{\partial }{\partial x} \left( \frac{\partial^2 q_1 }{ \partial x^2 }
 +3 \frac{\partial q_{2}}{\partial x}q_{3}  +3q_{1}q_{2}^{2} +q_{3}^{3} \right), \\
 \alpha \frac{\partial q_{2}}{\partial t} &=\frac{1}{4}\frac{\partial }{\partial x} \left(  -\frac{\partial^2 q_2 }{ \partial x^2 }
+3\frac{\partial}{\partial x}\left( q_{1}^{2}  -q_{3}^{2}  \right) +12 q_{1}q_{2}q_{3}- 2 q_{2}^{3}\right), \\
 \alpha \frac{\partial q_{3}}{\partial t} &=\frac{1}{2}\frac{\partial }{\partial x} \left( \frac{\partial^2 q_3 }{ \partial x^2 }
 -3 \frac{\partial q_{2}}{\partial x}q_{1}  +3q_{3}q_{2}^{2} +q_{1}^{3} \right).
\end{split}
\end{equation}
The corresponding Hamiltonian is
\begin{equation}
\begin{aligned}\label{eq:H1}
H&=\frac{1}{\alpha} \int_{-\infty}^{\infty} dx\;\biggl( \frac{1}{4}q_{1}^{4}-\frac{1}{8}q_{2}^{4}+\frac{1}{4}q_{3}^{4}+\frac{3}{2}q_{1}q_{2}^{2}q_{3}
+\frac{1}{2}q_{1}q_{2}\frac{\partial q_{1}}{\partial x}-\frac{1}{2}q_{1}^{2}\frac{\partial q_{2}}{\partial x}\\
&+\frac{1}{2}q_{3}^{2}\frac{\partial q_{2}}{\partial x}
-\frac{1}{6} \biggl( \frac{\partial q_{1}}{\partial x} \biggr) \biggl( \frac{\partial q_{3}}{\partial x} \biggr) +\frac{1}{24}
\biggl( \frac{\partial q_{2}}{\partial x} \biggr) ^{2}-\frac{1}{2}q_{2}q_{3}\frac{\partial q_{3}}{\partial x}\\
&+\frac{1}{6} q_{3}\frac{\partial^{2} q_{1}}{\partial x^{2}}-\frac{1}{12} q_{2}\frac{\partial^{2} q_{2}}{\partial x^{2}}+\frac{1}{6} q_{1}\frac{\partial^{2} q_{3}}{\partial x^{2}}
\biggr) .
\end{aligned}
\end{equation}
The next example is related to $A_4^{(1)}$.
The potential of the Lax operator now is
\begin{equation}
\label{eq:A4L}
\begin{split}
U(x,t,\lambda) = \left(\begin{array}{ccccc}
0 & q_1 & q_2 & q_3 & q_4  \\
q_4 & 0 & q_1 & q_2 & q_3  \\
q_3 & q_4 & 0 & q_1 & q_2  \\
q_2 & q_3 & q_4 & 0 & q_1  \\
q_1 & q_2 & q_3 & q_4 & 0
 \end{array}\right)
 - \lambda \left(\begin{array}{ccccc}
 1 & 0 & 0 & 0 & 0  \\
 0 & \omega & 0 & 0 & 0  \\
 0 & 0 & \omega^2 & 0 & 0  \\
 0 & 0 & 0 & \omega^3 & 0  \\
 0 & 0 & 0 & 0 & \omega^4
 \end{array}\right),
\end{split}
\end{equation}
where $\omega= e^{2\pi i/5}$.  The set of equations is
\begin{equation}\label{eq:A4mkdv}\begin{split}
 \alpha \frac{\partial q_{1}}{\partial t} &=\frac{\partial }{\partial x} \left( \frac{ c_1}{2s_1^2} \frac{\partial^2 q_1 }{ \partial x^2 }
   + \frac{3}{2s_1} q_4 \frac{\partial q_2}{ \partial x } + \frac{3}{2s_2} q_3 \frac{\partial q_3}{ \partial x }
  +  3 q_1 q_2 q_3 + q_2^3 + 3 q_3 q_4^2 \right),\\
 \alpha \frac{\partial q_{2}}{\partial t} &=\frac{\partial }{\partial x} \left( -\frac{ c_2}{2s_2^2} \frac{\partial^2 q_2 }{ \partial x^2 }
  - \frac{3}{2s_2} q_3 \frac{\partial q_4}{ \partial x } + \frac{3}{2s_1} q_1 \frac{\partial q_1}{ \partial x }
 +  3q_1 q_2 q_4 + q_4^3 + 3 q_1 q_3^2 \right),\\
 \alpha \frac{\partial q_{3}}{\partial t} &=\frac{\partial }{\partial x} \left( -\frac{ c_2}{2s_2^2} \frac{\partial^2 q_3 }{ \partial x^2 }
 + \frac{3}{2s_2} q_2 \frac{\partial q_1}{ \partial x } - \frac{3}{2s_1} q_4 \frac{\partial q_4}{ \partial x }
  +  3 q_1 q_3 q_4 + q_1^3 + 3 q_4 q_2^2 \right),\\
 \alpha \frac{\partial q_{4}}{\partial t} &=\frac{\partial }{\partial x} \left( \frac{ c_1}{2s_1^2} \frac{\partial^2 q_4 }{ \partial x^2 }
 - \frac{3}{2s_1} q_1 \frac{\partial q_3}{ \partial x } - \frac{3}{2s_2} q_2 \frac{\partial q_2}{ \partial x }
 +  3 q_2 q_3 q_4 + q_3^3 + 3 q_2 q_1^2 \right),
\end{split}\end{equation}
where
\begin{equation}\label{eq:sk}\begin{aligned}
s_k &= \sin \left( \frac{k\pi }{5} \right), &\quad c_k &= \cos \left( \frac{k\pi }{5} \right),  &\quad s_1 &= \frac{1}{4} \sqrt{10-2\sqrt{5}},\\
c_1 &= \frac{1}{4}(1+\sqrt{5}), &\quad s_2 &= \frac{1}{4} \sqrt{10-2\sqrt{5}}, &\quad c_2 &= \frac{1}{4}(\sqrt{5}-1).
\end{aligned}\end{equation}
The Hamiltonian is
\begin{equation}
\label{eq:HA4}\begin{aligned}
H &= \frac{2b}{3a^3} \int_{-\infty}^{\infty} dx\; \Bigg( - \frac{c_1}{2s_1^2} \frac{\partial q_1}{ \partial x }\frac{\partial q_4}{ \partial x }
+\frac{ c_2}{2s_2^2}  \frac{\partial q_2}{ \partial x } \frac{\partial q_3}{ \partial x } + q_1q_3^3 +q_2^3q_4 + q_3q_4^3 \\
&+ \frac{3}{8s_1} \left( q_4^2  \frac{\partial q_2}{ \partial x } -
2 q_2q_4  \frac{\partial q_4}{ \partial x } + 2q_1q_3  \frac{\partial q_1}{ \partial x } -  q_1^2  \frac{\partial q_3}{ \partial x }\right) + 3 q_1q_2q_3q_4 + q_1^3q_2   \\
&+ \frac{3}{8s_2} \left( q_2^2  \frac{\partial q_1}{ \partial x } -
2 q_1q_2  \frac{\partial q_2}{ \partial x } + 2q_3q_4  \frac{\partial q_3}{ \partial x } -  q_3^2  \frac{\partial q_4}{ \partial x }\right) \Bigg).
\end{aligned}
\end{equation}

The last example is related to $A_5^{(1)}$.
The potential of the Lax operator now is
\begin{eqnarray}
\label{eq:A5L}
U(x,t,\lambda) = \left(
\begin{matrix}
 0   & q_1 & q_2 & q_3 & q_4 & q_5 \\
 q_5 & 0   & q_1 & q_2 & q_3 & q_4 \\
 q_4 & q_5 & 0   & q_1 & q_2 & q_3 \\
 q_3 & q_4 & q_5 & 0   & q_1 & q_2 \\
 q_2 & q_3 & q_4 & q_5 & 0   & q_1 \\
 q_1 & q_2 & q_3 & q_4 & q_5 & 0   \\
\end{matrix}
 \right) - \lambda \left(
 \begin{matrix}
  1 & 0      & 0        & 0        & 0        & 0        \\
  0 & \omega & 0        & 0        & 0        & 0        \\
  0 & 0      & \omega^2 & 0        & 0        & 0        \\
  0 & 0      & 0        & \omega^3 & 0        & 0        \\
  0 & 0      & 0        & 0        & \omega^4 & 0        \\
  0 & 0      & 0        & 0        & 0        & \omega^5 \\
  \end{matrix}
  \right),
\end{eqnarray}
where $\omega= e^{\pi i/3}$.
Skipping the details we write down the corresponding equations
\begin{equation}\label{eq:A5eq}\begin{split}
\alpha \frac{\partial q_1 }{ \partial t } &= \frac{\partial }{ \partial x } \left( 4 \frac{\partial^2 q_1}{ \partial x^2 }
+ \sqrt{3} \left( 4 \frac{\partial q_2}{ \partial x } q_5 + 2q_3 \frac{\partial q_4}{ \partial x } + 3 \frac{\partial q_3}{ \partial x }q_4 \right) \right. \\
&+ 6q_3 (q_2^2 +q_5^2) + 3 q_1q_3^2 + 6q_4 (q_1q_2 + q_4q_5) \bigg), \\
\alpha \frac{\partial q_2 }{ \partial t } &= \frac{\partial }{\partial x} \left( \sqrt{3} \left( 4 \frac{\partial q_1}{ \partial x } q_1 + q_5 \frac{\partial q_3}{ \partial x } -
2 \frac{\partial q_5}{ \partial x }q_3 \right) 
+ 6q_1 (2 q_3q_4 + q_2q_5) + 3 q_2q_3^2 + 6q_4 q_5^2) \right) \\ 
\alpha \frac{\partial q_3 }{ \partial t } &= \frac{\partial }{ \partial x } \left( - \frac{\partial^2 q_3}{ \partial x^2 }
+ 2\sqrt{3} \left(  \frac{\partial q_2}{ \partial x } q_1 + 3q_2 \frac{\partial q_1}{ \partial x } -  \frac{\partial q_4}{ \partial x }q_5
- 3 \frac{\partial q_5}{ \partial x }q_4 \right) \right.  \\
&+ 12 (q_3 (q_2 q_4 + q_1q_5) +  q_1q_4^2 + q_2^2q_5)+  4q_1^3 -2q_3^3 +4q_5^3 \bigg), \\
\alpha \frac{\partial q_4 }{ \partial t } &= \frac{\partial }{\partial x} \left(  \sqrt{3} \left( 2 \frac{\partial q_1}{ \partial x } q_3 - q_1 \frac{\partial q_3}{ \partial x } -
4 \frac{\partial q_5}{ \partial x }q_5 \right)  
+ 6q_1 ( q_1q_2 + q_4q_5) + 3 q_4q_3^2 + 12q_2q_3 q_5 \right) \\ 
\alpha \frac{\partial q_5 }{ \partial t } &= \frac{\partial }{ \partial x } \left( -4 \frac{\partial^2 q_5}{ \partial x^2 }
+ \sqrt{3} \left( 4 \frac{\partial q_4}{ \partial x } q_1 + 2q_3 \frac{\partial q_2}{ \partial x } + 3 \frac{\partial q_3}{ \partial x }q_2 \right) \right. \\
&- 6q_3 (q_1^2 +q_4^2) - 3 q_5q_3^2 - 6q_2 (q_1q_2 + q_4q_5) \bigg).
\end{split}\end{equation}

The corresponding Hamiltonian is
\begin{equation}\label{eq:A5H}\begin{aligned}
H &= \frac{2b}{3a^3} \int_{-\infty}^{\infty} dx\; \left( 3 \frac{\partial q_1}{ \partial x }\frac{\partial q_5}{ \partial x } -\frac{ 3}{16} \left( \frac{\partial q_3}{ \partial x }\right)^2 - \sqrt{3} \left( \frac{5}{4}q_2q_3 + q_1q_4\right) \frac{\partial q_1}{ \partial x } \right. \\
 &+ \sqrt{3} \left( \frac{1}{4}q_1q_3 - q_5^2\right) \frac{\partial q_2}{ \partial x } +\sqrt{3} \left( q_1q_2 - q_4q_5\right) \frac{\partial q_3}{ \partial x }
 + \sqrt{3} \left( q_1^2 -\frac{1}{4}q_3q_5 \right) \frac{\partial q_4}{ \partial x } \\
&+\sqrt{3} \left( q_2 q_5 +\frac{5}{4}q_3q_4 \right) \frac{\partial q_5}{ \partial x } - \frac{9}{4} (q_1q_5 +q_2q_4)q_3^2
- \frac{9}{4} (q_1q_2 +q_4q_5)^2 \\
&- \frac{3}{2} (q_1^3 +q_5^3)q_3 - \frac{9}{2} (q_2^2q_5 +q_1q_4^2)q_3 \bigg).
\end{aligned}
\end{equation}

\section{Additional Involutions }

Along with the $\bbbz_{r+1}$-reduction (\ref{2}) we can introduce one of the following  involutions
($\bbbz_2$-reductions) on the Lax pair:
\begin{equation}\label{eq:14}
\begin{aligned}
&\mbox{a)} &\quad K_0^{-1} U^\dag (x,t,\kappa_1(\lambda) ) K_0 &=  U(x,t,\lambda ) , &\quad \kappa_1(\lambda)&=\omega^{-1}\lambda^*; \\
&\mbox{b)} &\quad K_0^{-1} U^* (x,t,\kappa_1(\lambda)) K_0 &=  -U(x,t,\lambda ) , &\quad \kappa_1(\lambda) &=-\omega^{-1}\lambda^*;\\
&\mbox{c)} &\quad  U^T (x,t,-\lambda )&= -  U(x,t,\lambda ) ,
\end{aligned}
\end{equation}
where $K_0^2=\openone$.
If we choose
\begin{equation}
K_0 = \sum_{k=1}^{r+1} E_{k,r-k+2} \,
\end{equation}
then the action of $K_0$ on the basis is as follows
\begin{equation}\label{eq:K0}
\begin{aligned}
K_0 \left( J_s^{(k)}\right)^\dag K_0 &= \omega^{k(s-1)} J_s^{(k)}, &\quad K_0 \left( J_s^{(k)}\right)^* K_0 &= \omega^{-k} J_{-s}^{(k)}.
\end{aligned}
\end{equation}

An immediate consequences of equation (\ref{eq:14}) are the constraints on the potentials
\begin{equation}\label{eq:14U}
\begin{aligned}
&\mbox{a)} &\quad K_0^{-1} Q^\dag (x,t) K_0 &= Q(x,t), &\qquad  K_0^{-1} (J_0^{(1)})^\dag  K_0 &= \omega^{-1} J_0^{(1)} ,\\
&\mbox{b)} &\quad K_0^{-1} Q^* (x,t) K_0 &= -Q(x,t), &\qquad  K_0^{-1} (J_0^{(1)})^* K_0 &= \omega^{-1} J_0^{(1)} ,\\
&\mbox{c)} &\quad  Q^T (x,t) &= -Q(x,t), &\qquad (J_0^{(1)})^T &= J_0^{(1)}.
\end{aligned}
\end{equation}
More specifically from equation (\ref{eq:14U}) follows that each of the algebraic relations below
\begin{equation}
\label{140}
\begin{aligned}
&\mbox{a)} &\quad q_{j} ^* (x,t) &=  q_{j} (x,t), &\qquad  \alpha&=\alpha^*;\\
&\mbox{b)} &\quad q_{j} ^* (x,t) &=  -q_{r-j+1} (x,t), &\qquad \alpha&=\alpha^*; \\
&\mbox{c)} &\quad q_{j}  (x,t) &= -  q_{r-j+1} (x,t),
\end{aligned}
\end{equation}
where $j=1,\dots, r$, are compatible with the evolution of the mKdV equations (\ref{eq:30'}).


These involutions reduce the form of the scattering matrix
($\bbbz_2$-reductions)
\begin{equation}
\begin{aligned}
&\mbox{a)} &\quad K_{0}^{-1} T^{\dag} (\kappa_1(\lambda),t ) K_{0} &=  T^{-1}(\lambda,t ), \\
&\mbox{b)} &\quad K_{0}^{-1} T^{*} (\kappa_1(\lambda),t) K_{0} &=  T(\lambda,t ), \\
&\mbox{c)} &\quad  T^{T} (-\lambda,t )&=  T^{-1}(\lambda,t ) .
\end{aligned}
\end{equation}

If we apply case a) of (\ref{140}) we get the same set of mKdV equations
with $q_{1}, q_{2}$ and $q_{3}$ being purely real functions.
In the case b) we put $q_{1}=-q_{3}^*=u$ and $q_{2}=-q_{2}^*=iv$ and we get
\begin{equation}
\label{eq:mKdVb}
\begin{split}
\alpha \frac{\partial v}{\partial t} &=- \frac{1}{4}\frac{\partial^{3} v}{\partial x^{3}}+\frac{3}{4i}\frac{\partial^{2}}{\partial x^{2}}\left( u^{2} - (u^{*})^2 \right) - 3 \frac{\partial}{\partial x}({|u|}^{2}v)+\frac{1}{2}\frac{\partial}{\partial x}v^{3},\\
\alpha \frac{\partial u}{\partial t}& =\frac{1}{2}\frac{\partial^{3} u}{\partial x^{3}}-i \frac{3}{2}\frac{\partial}{\partial x}\left(u^{*} \frac{\partial v}{\partial x} \right)-\frac{3}{2}\frac{\partial}{\partial x}(uv^{2})-\frac{\partial}{\partial x}(u^{*})^{3},
\end{split}
\end{equation}
where $u$ is a complex function but $v$ is a purely real function.
The corresponding Hamiltonian is
\begin{equation}\label{eq:}\begin{split}
H & =\frac{1}{\alpha} \biggl( 
\frac{1}{4}u^{4}-\frac{1}{8}v^{4}+\frac{1}{4}(u^{*})^{4}+\frac{3}{2}|u|^{2}v^{2}
+\frac{i}{2}uv\frac{\partial u}{\partial 
x}-\frac{i}{2}u^{2}\frac{\partial v}{\partial x} 
+\frac{i}{2}(u^{*})^{2}\frac{\partial v}{\partial x} \\
& \qquad +\frac{1}{6} \biggl| \frac{\partial u}{\partial x} \biggr|^{2} 
-\frac{1}{24}
\biggl( \frac{\partial v}{\partial x} \biggr) 
^{2}-\frac{i}{2}u^{*}v\frac{\partial u^{*}}{\partial x}
  -\frac{1}{6} u^{*}\frac{\partial^{2} u}{\partial x^{2}}+\frac{1}{12} 
v\frac{\partial^{2} v}{\partial x^{2}}-\frac{1}{6} u\frac{\partial^{2} 
u^{*}}{\partial x^{2}}
\biggr).
\end{split}\end{equation}
The case \mbox{c)} leads to the well known defocusing mKdV equation
\begin{eqnarray}
\alpha \frac{\partial u}{\partial t}=\frac{1}{2}\frac{\partial^{3} u}{\partial x^{3}}-\frac{\partial}{\partial x}(u^{3}),
\end{eqnarray}
where $u$ is a complex function.
The corresponding Hamiltonian is
\begin{equation}
H=-\frac{1}{4\alpha} \left( \left( \frac{\partial u}{\partial x} \right)^{2}+u^{4} \right).
\end{equation}

And finally, considering $A^{(1)}_{5}$ algebra with $\bbbd_6$-reduction, case \mbox{c)} we find
\begin{equation}\label{eq:00}
\begin{split}
\alpha \frac{\partial u}{\partial t} &=2 \frac{\partial^{3} u}{\partial x^{3}}-2\sqrt{3}\frac{\partial}{\partial x}
\left(u \frac{\partial v}{\partial x} \right)-6\frac{\partial}{\partial x}(uv^{2}),\\
\alpha \frac{\partial v}{\partial t} &=\sqrt{3}\frac{\partial^{2}}{\partial x^{2}}\left(u^{2} \right)-6\frac{\partial}{\partial x}(u^{2}v),
\end{split}
\end{equation}
where $u$ and $v$ are complex functions.
The Hamiltonian is given by
\begin{equation}
H=-\frac{1}{\alpha} \left( \left( \frac{\partial u}{\partial x} \right)^{2}+\sqrt{3}u^{2} \left( \frac{\partial v}{\partial x} \right) +3u^{2}v^{2} \right).
\end{equation}

\section{Generating the $M$-operators of NLEE}

Here $\chi_{\nu}(x,\lambda)$ is a fundamental analytic solution (FAS), analytic in the sector
$\Omega_{\nu}$ of the complex plane. Let us introduce
\begin{equation}
\label{functional}
\mathcal{K}^{(m)}(x,t,\lambda)=\lambda^{m} \chi_{\nu} J^{m} \hat{\chi}_{\nu} (x,t,\lambda),
\end{equation}
which is the generating functional of the $M$-operators for the mKdV equations.
In this representation $J^m$ is a constant diagonal matrix and $\chi_{\nu}(x,\lambda)$
is a FAS of the Lax operator. From (\ref{functional}) and using
\begin{equation}
i\frac{\partial \chi_{\nu}}{\partial x}= -(Q-\lambda J)\chi_{\nu}
\end{equation}
we find that
\begin{equation}
i \frac{\partial \mathcal{K}^{(m)}}{\partial x} + [Q-\lambda J,\mathcal{K}^{(m)}]=0,
\end{equation}
where we can use the following form
\begin{equation}
\label{asymptotic}
\mathcal{K}^{(m)}(x,t,\lambda) = \lambda^{m}\left( J^m +\sum_{s=1}^{\infty} \lambda^{-s} \mathcal{K}_{s}^{(m)}(x,t) \right) \,.
\end{equation}
We can view $\chi$ as FAS of $L$ admitting asymptotic expansion for details see \cite{PLA110,SIAM*14}
\begin{equation}
\chi = \openone +\sum_{s=1}^{\infty} \lambda^{-s} \chi_{s} \,.
\end{equation}
This will allow us to calculate the first few expansion coefficients, which are integrals of motion for the mKdV equations.
From (\ref{asymptotic}) we find for $s$ from $1$ to infinity
\begin{equation}
i \frac{\partial \mathcal{K}_{s}^{(m)}}{\partial x}+ [Q,\mathcal{K}_{s}]=[J,\mathcal{K}_{s+1}^{(m)}]
\end{equation}
and for $s=0$
\begin{equation}
[Q,J^m]=[J,\mathcal{K}_{1}^{(m)}]
\end{equation}
which coincide with the recursion relations for the $M$ operator (\ref{Recurrence1}) if we replace $s$ with $(m-s)$. Solving them we get
\begin{equation}
\mathcal{K}_{1}^{m,\bot}=\mbox{ad}_{J}^{-1}[Q,J^m]
\end{equation}
\begin{equation}
\mathcal{K}_{s+1}^{m,\bot}=\mbox{ad}_{J}^{-1}\biggl( i\frac{\partial \mathcal{K}_{s}^{(m)}}{\partial x}+[Q,\mathcal{K}_{s}^{(m)}] \biggr)
\end{equation}
\begin{equation}
\mathcal{K}_{s}^{m,\|}= \frac{i}{r+1} (\partial_x)^{-1} \left< \left[Q(x,t),\mathcal{K}_{s}^{m,\bot}\right], J^{r+1-s} \right>J^{s}
\end{equation}
or
\begin{align*}
&\mathcal{K}^{(m)}=\lambda^{m} J^m+\lambda^{m-1}\mbox{ad}_{J}^{-1}[Q,J^m]+
\sum_{s=2}^{\infty} \lambda^{m-s} \mbox{ad}_{J}^{-1}\biggl( i\frac{\partial \mathcal{K}_{s-1}^{(m)}}{\partial x}+[Q,\mathcal{K}_{s-1}^{(m)}] \biggr)+\\
&+ \sum_{s=1}^{\infty} \lambda^{m-s} \frac{i}{r+1} (\partial_x)^{-1} \left< \left[Q(x,t),\mathcal{K}_{s}^{m,\bot}\right], J^{r+1-s} \right>J^{s}.
\end{align*}

Here we can also consider $s_{a}= a(r+1)+s$. Then
\[ K_{s_a}^{(m)} = \boldsymbol{\Lambda}^a K_s^{(m)}, \qquad \boldsymbol{\Lambda} = \Lambda_0\Lambda_1 \cdots \Lambda_{r}. \]
Thus we can express the $M$ operators and the whole hierarchy of NLEE in compact form through the recursion operators
\cite{Dickey}. For the general case when $M$ is polynomial from $m$-th order we have
\begin{equation}
M^{(m)} (x,t , \lambda) =(\mathcal{K}^{(m)}(x,t,\lambda))_{+}.
\end{equation}

\subsection{The integrals of motion}

We can use the generating functional $\mathcal{K}_{s}(x,t,\lambda)$ also to derive the integrals of motion for the mKdV type equations
derived above. To this end we have to consider the resulting functionals
\begin{equation}
\label{eq:IM1}
\begin{split}
\mathcal{D}_{s,m} (\lambda) = \left\langle \mathcal{K}^{(m)}(x,t,\lambda), J_0^{r+1-s} \right\rangle.
\end{split}
\end{equation}

First, $\mathcal{D}_{s,m} (\lambda)$ are independent of time.
Second, $\mathcal{D}_{s,m} (\lambda)$ allow asymptotic expansion over the inverse powers of $\lambda$.
Indeed using (\ref{asymptotic}) we find that
\begin{equation}
\label{eq:Dm}
\mathcal{D}_{s, m} (\lambda) = \sum_{k=1}^{\infty} \mathcal{D}_{s,m}^{(k)} \lambda^{m-k} = \sum_{p=1}^{\infty} \mathcal{D}_{s, m}^{((r+1)p)} \lambda^{m-(r+1)p}.
\end{equation}

Using the properties of the Killing form, namely that
\begin{equation}\label{eq:Kill}\begin{split}
\langle X^{(m)},  Y^{(s)} \rangle \simeq \delta_{m+s}^{(r+1)},
\end{split}\end{equation}
where $\delta_{m+s}^{(r+1)} =1$ only if $(m+s)=0 \mbox{ mod } (r+1)$ we find that
there are gaps in the sequences of the integrals of motion \cite{Mikh}. Indeed, only the $\mathcal{D}_{s,m}^{(k)}$ for which
$k$ is proportional to $r+1$ are nontrivial; all the others vanish.

\section{Discussion and conclusions}

In the present paper we have derived the recursion operators of the hierarchy of soliton equations related to the affine Lie algebras $A^{(1)}_{r}$.
These equations belong to the hierarchy containing the two-dimensional Toda field theories discovered by
Mikhailov \cite{Mikh}. We have imposed a $\bbbz_{r+1}$-reduction on the corresponding Lax operator $L$.
We have also demonstrated several examples that are obtained from the generic mKdV equations by imposing additional  $\bbbz_2$-reductions.

A natural extension of these results involve  the construction of their soliton solutions using the dressing method
\cite{Mikh,ZaSha1,ZaSha2,ZaMi,MiZa*80} and analysis of their properties. The fundamental analytic solutions for this class
of Lax operators have been constructed in \cite{GYa*94} and the effects of the $\mathbb{Z}_h$ reductions, where $h$
is the Coxeter number
were studied in \cite{VG-Ya-13,SIAM*14}. Still more detailed study of their spectral properties and of the expansions over
their squared solutions could lead to the construction of their Hamiltonian hierarchies and in particular to the derivation of the
action-angle variables for these NLEE. These and other aspects of the theory of soliton equations will be published elsewhere.

\section*{Acknowledgements}
The work is supported in part by the ICTP - SEENET-MTP project PRJ-09. One of us (VSG) is grateful to professor A. S. Sorin for
useful discussions during his visit to JINR, Dubna, Russia under project 01-3-1116-2014/2018.

\bibliographystyle{amsplain}

\begin{thebibliography}{99}

\bibitem{GGKM}
C.S. Gardner, J.M. Greene, M.D. Kruskal, and R.M. Miura,
"Method for solving the Korteweg-de Vries equation",
Phys. Rev. Lett. 19, 10951097 (1967).

\bibitem{Lax}
P.D. Lax,
"Integrals of nonlinear equations of evolution and solitary waves",
Comm. Pure Appl. Math. {\bf 21}, 467-490 (1968).

\bibitem{ZaSha1}
V.E. Zakharov, A.B. Shabat,
"A scheme for integrating the nonlinear equations of mathematical physics by the method of the inverse scattering problem. I.",
Funct. Anal. Appl. {\bf 8}, 226-235 (1974).

\bibitem{ZaFa}
V.E. Zakharov, L.D. Faddeev,
"Korteweg-de Vries equation: A completely integrable Hamiltonian system",
Funct. Anal.  Appl. {\bf 8}, 226-235 (1974).

\bibitem{ZaSha2}
V.E. Zakharov, A.B. Shabat,
{\it Integration of nonlinear equations of mathematical physics by the method of inverse scattering. II.},
Funct. Anal. Appl. {\bf 13}, 166-174 (1979).

\bibitem{CalDeg}
F. Calogero, A. Degasperis,
{\it Spectral Transform and Solitons}, Vol. I. (North Holland, Amsterdam, 1982).

\bibitem{FaTa}
L.D. Faddeev, L.A. Takhtadjan,
{\it Hamiltonian Methods in the Theory of Solitons},
(Springer Verlag, Berlin, 1987).

\bibitem{GYaV*08}
V.S. Gerdjikov, G. Vilasi, A.B. Yanovski,
{\it Integrable Hamiltonian Hierarchies. Spectral and Geometric Methods},
Lecture Notes in Physics \textbf{748}, (Springer, Berlin, Heidelberg, New York, 2008).

\bibitem{ZMNP}
S.P. Novikov, S.V. Manakov, L.P. Pitaevskii, and V.E. Zakharov,
{\it Theory of Solitons: The Inverse Scattering Method},
(Plenum, Consultants Bureau, New York, 1984).

\bibitem{AKNS}
M.J. Ablowitz, D.J. Kaup, A.C. Newell, H. Segur,
"The inverse scattering transform --- Fourier analysis for nonlinear problems",
Studies in Appl. Math. \textbf{53}, 249-315 (1974).

\bibitem{GeKh1}
V.S. Gerdjikov, E.Kh. Khristov,
"On the evolution equations solvable with the inverse scattering problem. I. The spectral theory",
Bulgarian J. Phys. {\bf 7,} No.1, 28--41 (1980) (In Russian); \\
%
 V.S. Gerdjikov, E.Kh. Khristov,
 "On the evolution equations solvable with the inverse scattering problem. II. Hamiltonian structures and Backlund transformations",
Bulgarian J. Phys. {\bf 7,} No.2, 119--133 (1980)  (In Russian).

\bibitem{Kaup}
D.J. Kaup,
"Closure of the Squared Zakharov-Shabat Eigenstates",
J. Math. Analysis and Applications {\bf  54}, 849-864 (1976).

\bibitem{Ge4}
V.S. Gerdjikov,
"Generalised Fourier transforms for the soliton equations. Gauge covariant formulation",
Inverse Problems \textbf{2}, 51-74 (1986).

\bibitem{GeKu*81}
V.S. Gerdjikov, P.P. Kulish,
"The generating operator for the $n \times n$ linear system",
Physica  {\bf 3D}, 549-564 (1981).

\bibitem{SIAM*14}
V.S. Gerdjikov, A.B. Yanovski,
"CBC systems with Mikhailov reductions by Coxeter Automorphism: I. Spectral Theory of the Recursion Operators",
Stud. Appl. Math. (2014) (In press).

\bibitem{Mikh}
A.V. Mikhailov,
"The reduction problem and the inverse scattering problem",
Physica \textbf{3D}, 73-117 (1981).

\bibitem{DriSok}
V.V. Drinfel'd, V.G. Sokolov,
"Lie algebras and equations of Korteweg-de Vries type",
Itogi Nauk. i Techn., Seriya Sovremennye Problemy Matematiki (Noveishie Dostizheniya),
\textbf{24}, 81-180 (1984).

\bibitem{GMSV1}
V.S. Gerdjikov, D.M. Mladenov, A.A. Stefanov, and S.K. Varbev,
"MKdV-type of equations related to $\mathfrak{sl}(N, \mathbb{C})$ algebra",
(Cambridge Scholar Publishing, 2014) (In press).

\bibitem{GMSV2}
V.S. Gerdjikov, D.M. Mladenov, A.A. Stefanov, S.K. Varbev,
"On a one-parameter family of mKdV equations related to the $\mathfrak{so}(8)$ Lie algebra",
(Cambridge Scholar Publishing, 2014) (In press).

\bibitem{GMSV3}
V.S. Gerdjikov, D.M. Mladenov, A.A. Stefanov, S.K. Varbev,
"MKdV-type of equations related to $B_{2}^{(1)}$ and $A_{4}^{(2)}$ algebra",
(Springer, Berlin, Heidelberg, New York, 2014) (In press).

\bibitem{67}
V.S. Gerdjikov,
"Algebraic and analytic aspects of $N$-wave type equations",
Contemporary Mathematics \textbf{301}, 35-68 (2002).

\bibitem{VSG-88}
V.S. Gerdjikov,
"$Z_{N}$-reductions and new integrable versions of derivative nonlinear Schr\"o\-dinger equations",
{\it Nonlinear Evolution Equations: Integrability and Spectral Methods},
Ed. A.P. Fordy, A. Degasperis, M. Lakshmanan,
(Manchester University Press, 1988), pp. 367-372.

\bibitem{VG-13}
V.S. Gerdjikov,
"Derivative nonlinear Schr\"odinger equations with $\mathbb{Z}_{N}$ and $\mathbb{D}_{N}$ reductions",
Romanian J. Phys. \textbf{58}, 573-582 (2013).

\bibitem{KauNew}
D.J. Kaup, A.C. Newell,
"Soliton equations, singular dispersion relations and moving eigenvalues",
Adv. Math. \textbf {31}, 67-100 (1979).

\bibitem{VG-Ya-13}
V.S. Gerdjikov, A.B. Yanovski,
"On soliton equations  with $\mathbb{Z}_{ {h}}$ and $\mathbb{D}_{{h}}$ reductions: conservation laws and generating operators",
J. Geom. Symm. Phys. \textbf{31}, 57-92 (2013).

\bibitem{Yan2013a}
A.B. Yanovski,
"Recursion operators and expansions over adjoint solutions for the Caudrey-Beals-Coifman system with $\mathbb{Z}_p$ reductions of Mikhailov type",
J. Geom. Symm. Phys. \textbf{30}, 105-119 (2013).

\bibitem{YanVi*2012b}
A.B. Yanovski, G. Vilasi,
"Geometric theory of the recursion operators for the generalized Zakharov-Shabat system in pole gauge on the algebra $sl(n, \mathbb{C})$: with and without reductions",
SIGMA \textbf{8}, 87-110 (2012).

\bibitem{Cart}
R. Carter,
{\it Lie Algebras of Finite and Affine Type}
(Cambridge University Press, Cambridge, 2005).

\bibitem{Helg}
S. Helgasson,
{\it Differential geometry, Lie groups and symmetric spaces},
(Academic Press, New York, 1978).

\bibitem{Kac}
V. Kac,
{\it Infinite-Dimensional Lie Algebras},
(Cambridge University Press, Cambridge, 1994).

\bibitem{PLA110} V. S. Gerdjikov, A. B. Yanovsky,\
{\it Gauge covariant formulation  of  the  generating operator. II. Systems on homogeneous spaces},\
Phys. Lett. A, {\bf 110A,} n. 2, 53-58, (1985).

\bibitem{Dickey}
L.A. Dickey,
{\it Soliton Equations and Hamiltonian Systems}
(World Scientific, Singapore, 1991).

\bibitem{MiZa*80}
A.V. Mikhailov, V.E. Zakharov,
"On the integrability of classical spinor models in two-dimensional space-time",
Comm. Math. Phys. \textbf {74},  21-40 (1980).

\bibitem{ZaMi}
V.E. Zakharov, A.V. Mikhailov,
"Relativistically invariant two-dimensional models of field theory which are integrable by means of the inverse scattering problem method",
Soviet Phys. JETP \textbf {47}, 1017-1027 (1978).

\bibitem{GYa*94}
V.S. Gerdjikov, A.B. Yanovski,
"Completeness of the eigenfunctions for the Caudrey-Beals-Coifman system",
J. Math. Phys. \textbf {35}, 3687-3725 (1994).

\end{thebibliography}

\end{document}